\documentstyle[epsfig,12pt]{article}
\let\section=\subsection     \let\subsection=\subsubsection

\newcommand{\be}{\begin{equation}}  
\newcommand{\ee}{\end{equation}}  
\newcommand{\bd}{\begin{displaymath}}  
\newcommand{\ed}{\end{displaymath}}  
\newcommand{\ba}{\begin{eqnarray}}  
\newcommand{\ea}{\end{eqnarray}}

\begin{document}
\begin{center}
 {\large\bf IN-MEDIUM  PION-PION INTERACTION}\\[2mm]
 {\large\bf AND CHIRAL SYMMETRY RESTORATION}\\[5mm]
D.  DAVESNE$^1$, Y. J. ZHANG$^2$, G. CHANFRAY$^1$, J. WAMBACH$^3$  \\[5mm] 
{\small\it $^1$ IPN Lyon, 43 Bd  du 11 Novembre  1918, 
F-69622 Villeurbanne C\'edex, France\\
$^2$ Joint Laboratory for Quantum Optics
    Shanghai Institute of Optics and Fine Mechanics, Academia Sinica
    P. O. Box 800-211, Shanghai 201800, China\\
$^3$  IKP, Technische Universit\"at    
Darmstadt, \\ Schlo{\ss}gartenstra{\ss}e 9, 64289 Darmstadt, Germany.\\[8mm]}
\end{center}

\begin{abstract}\noindent  
We discuss medium modifications of the unitarized pion-pion interaction in
the nuclear medium. We incorporate both the effects of chiral symmetry
restoration and the influence of collective nuclear pionic modes, 
originating from the p-wave coupling of the pion to delta-hole configurations. 
We show how the resulting strong enhancement of the sigma-meson spectral
function is related to large fluctuations of the condensate associated with the
partial restoration of chiral symmetry.

\end{abstract} 
\vspace{0.6cm}

\section{Introduction}

There are at least two excellent reasons to study the in-medium modifications
of the pion-pion interaction in the scalar-isoscalar channel both being 
related to fundamental questions in present day nuclear physics. 
The first refers the binding energy of nuclear matter since a 
modification of the correlated two-pion exchange may have
some deep consequences on the saturation mechanism. The second one 
is the direct connection with chiral symmetry restoration. From very general
arguments based on QCD and model calculations, partial chiral symmetry
restoration is expected to occur in nuclear matter. Hence, there must be a
softening of a collective scalar-isoscalar mode, usually called the sigma
meson, which becomes degenerate with its chiral partner {\it i.e.} the pion 
at full restoration density. This also implies that at some density 
the sigma-meson spectral function should exhibit a significant enhancement 
near the two-pion threshold. This effect can be seen as a precursor
of chiral symmetry restoration associated with large fluctuations 
of the quark condensate near the chiral phase transition \cite{Hat_99}.    

\noindent
However, the first proposed medium effect was the modification of the two-pion 
propagator and the unitarized $\pi\pi$ interaction from the
softening of the pion dispersion relation by p-wave coupling 
to $p-h$ and $\Delta-h$ states.
The existence of collective pionic modes produces a strong accumulation of 
strength near the two-pion threshold 
in the scalar-isoscalar channel \cite{Aou_94}. According to recent calculations
\cite{Sch_98,Kre_98}, this reshaping of the strength may provide a partial 
explanation  of the $\pi-2\pi$ 
data obtained  on various nuclei by the CHAOS collaboration 
at TRIUMF \cite{Bon_96}. These results
have been questioned in a recent paper  where it is found
that pion absorption forces the reaction to occur in the nuclear surface,
{\it i.e.} at very low density \cite{Vic_99}. It is clear that the effect 
of chiral symmetry restoration has to be included on top of p-wave pionic 
effects to get a better explanation of the data. One attempt 
to combine both effects is based on the linear sigma model (implemented 
with a form factor fitted to phase shifts) in which the sigma mass is dropped 
through a Brown-Rho scaling relation \cite{Aou_99}. 
It is found that chiral symmetry restoration 
increases the strength of the threshold enhancement by about a factor four 
as reported in these proceedings \cite{Aou_99}. Here we will discuss another attempt 
based on the Nambu-Jona-Lasinio model \cite{Da_99}. From this NJL model we derive an
in-medium pion-pion potential, formally equivalent to the linear sigma model, 
but with parameters ($m_\sigma, m_\pi, f_\pi$) replaced by their in-medium
values. Hence, at variance with a pure dropping sigma mass scenario, the basic
$\sigma\pi\pi$ and $4\pi$ couplings are also modified in a chirally consistent
framework. P-wave coupling of the pion is incorporated within a standard
nuclear-matter approach since the NJL model completely misses (at least in its present 
treatment) the phenomenologically well-established strong screening effect
from short-range correlations ($g'$ parameter). 
The underlying philosophy can be summarized in stating  
that the medium-modified soft physics linked to chiral symmetry 
($m_\pi, f_\pi$, low energy $\pi-\pi$ potential ) 
is calculated within the NJL model while p-wave physics yielding  
pionic nuclear collective modes is described through standard  
nuclear phenomenology.  
 
\section{ NJL and Density Dependent 
Linear Sigma Model}

We start with the $SU(2)$ version of the NJL model~:                 
${\cal L}= \bar{\psi}(i \partial-m)\psi 
         +g[(\bar{\psi}\psi)^2+(\bar{\psi}i\gamma_5\tau_j\psi)^2]$. 
At finite density, the gap equation which determines the constituent
quark $M$ mass reads~:
\begin{eqnarray}
M & = & m + 4N_cN_f g \int _{k_F} ^{\Lambda} \frac{k^2 dk}{2\pi ^2}\frac{M}{E}
\end{eqnarray} 
where $\Lambda$ is the (three-momentum) cutoff. The pion and the sigma meson 
can be constructed
within the standard RPA approximation. Limiting ourselves to zero momentum,
the meson propagators are obtained according to~:
\begin{eqnarray}
g^2_{\pi q q} D_\pi(\omega)&=&\bigg( \omega^2 I(\omega)-m_\pi^2 I(m_\pi)
\bigg)^{-1} \nonumber\\ 
g^2_{\sigma q q} D_\sigma(\omega)&=&\bigg( (\omega^2-4 M^2) I(\omega)-m_\pi^2 
I(m_\pi)\bigg)^{-1}  \nonumber\\
\hbox{with}\quad 
I(\omega)&=&2 N_c N_f\,\int _{k_F}^\Lambda {p^2 dp\over 2\pi^2}\,{1\over E_p\,
(4E^2_p-\omega^2)}
\end{eqnarray}  
The in-medium masses of the pion and the sigma meson are  
found as the pole of the above propagators. 
It is important to notice that they are pure collective 
$q \bar q$ states since, at zero momentum, there is no contribution 
from particle-hole excitations (Fermi sea). According to what is said above, the 
particle-hole sector is better treated within a standard nuclear physics approach. 
In addition, for reason of simplicity, we use in the following a 
simplified scheme where $I(\omega)$ is ``frozen'' at $\omega=0$.
We have verified numerically that this approximation gives almost 
the same result than the exact calculation for the pion mass, the sigma-meson 
mass and also for the pion decay constant (see \cite{Da_99}). 
The parameters of the model ($m, \Lambda, g$) have been adjusted
to obtain the vacuum values $f_\pi= 93\, MeV$, $m_\pi=139\, MeV$ and 
$m_\sigma=1\,GeV$  which is precisely the value of the bare sigma-meson mass, 
systematically used in our previous works \cite{Aou_94,Sch_98,Aou_99}. 
The s-wave optical potential is somewhat too small ($4\, MeV$ at $\rho_0$) 
but the incorporation of a phenonenologically more appropriate s-wave optical potential
($10\, MeV$) has only minor influences on the medium effects presented 
in this paper. 

\noindent 
We are now in position to construct vacuum and in-medium $\pi\pi$ 
potentials following the method developed in \cite{Qua_95} but adapted 
to finite density. The basic diagrams are box diagrams with four internal
quark lines for the direct $4\pi$ interaction and three internal quark 
lines for the $\pi\pi\sigma$ coupling from which one obtains the sigma exchange
diagram (Fig.~1). Keeping the full momentum dependence makes the problem of
subsequent unitarization hopelessly complex. As in \cite{Qua_95}, 
we first limit the calculation 
to the case where the four pions have momenta $p_i$ such that 
$p_i^2=m^2_\pi$ (what we call the exact scheme in \cite{Da_99}). However to make the
unitarization tractable we further simplify the calculation by freezing 
the momentum to $p=0$ in the quark-loop integral (simplified scheme). 
We have checked numerically that the resulting low-energy $\pi\pi$ amplitudes 
are almost identical in the simplified and exact schemes \cite{Da_99}. 
We obtain for the coupling constants
$\lambda_{4\pi}\equiv \lambda=(m^2_\sigma-m^2_\pi)/ 2 f^2_\pi$ and
$\lambda_{\sigma\pi\pi}=\lambda\, f_\pi$.
This is the well-known result of the linear sigma model but
with medium-modified parameters ($m_\sigma, m_\pi, f_\pi$). The
linear sigma model, seen as a $O(N+1)$ model with $N=3$, is treated to leading
order in the $1/N$ expansion which fulfills all the constraints (Ward identities)
of chiral symmetry \cite{Sch_98}. We supplement the model by adding a one-parameter 
form factor $v(k)$ to fit the phase shifts in vacuum once the scattering 
amplitude is unitarized.  The (in-medium) unitarized scalar-isoscalar 
$\pi\pi$ T matrix (in the CM frame   and for total  energy $E$
of the pion pair) is \cite{Sch_98,Da_99}~:     
\begin{equation}
\langle{\bf k}, -{\bf k}|T(E)|{\bf k}', -{\bf k}'\rangle=v(k) v(k')\,
{6 \lambda (E^2-m^2_\pi)\over 1-3\lambda \Sigma(E)}\, D_\sigma(E)
\end{equation}
where the unitarized sigma propagator ({\it i.e} with two-pion loop)  
$D_\sigma(E)$ is~:     
\begin{equation}
D_\sigma(E)=\bigg( E^2-m^2_\sigma- {6 \lambda^2 f^2_\pi \Sigma(E)\over 
1-3\lambda \Sigma(E)}\bigg)^{-1}\label{PROP}
\end{equation}
Chiral symmetry restoration is accounted for by the in-medium 
pion- as well as sigma-meson masses and $f_\pi$. The p-wave collective 
effects are embedded in the two-pion loop~:
\begin{equation}
\Sigma(E)=\int {d{\bf q}\over (2\pi)^3}\, v(q)\,
\int{i \,dq_0\over 2\pi}\, D_\pi({\bf q}, q_0)\, D_\pi(-{\bf q},\, E-q_0).
\end{equation}
The pion propagator $D_\pi({\bf q}, q_0)$ is calculated in a standard 
nuclear matter approach \cite{Da_99} and incorporates 
the p-wave coupling of the pion to
delta-hole states with short-range screening described by the usual
$g'_{\Delta\Delta}=0.5$ parameter. It is particularly interesting to inspect 
the sigma-meson spectral function. The result of the calculation is shown 
on Fig.~2. At twice normal nuclear-matter, which is close to the density 
where the quasi-pole in $D_\sigma$ is $E=2m_\pi$, we find a very sharp peak 
which can be understood as a precursor effect of chiral symmetry restoration. 
At normal nuclear-matter density the threshold peak is enhanced by a 
factor three as compared to a pure p-wave calculation. 
This is in qualitative
agreement with the dropping sigma-meson mass calculation where a factor four is
obtained \cite{Aou_99}. At density $0.5\, \rho_0$, more relevant for the CHAOS experiment,
we still have a sizable low-energy reshaping which could help to
explain the data.

\begin{figure}
    \centering\epsfig{figure=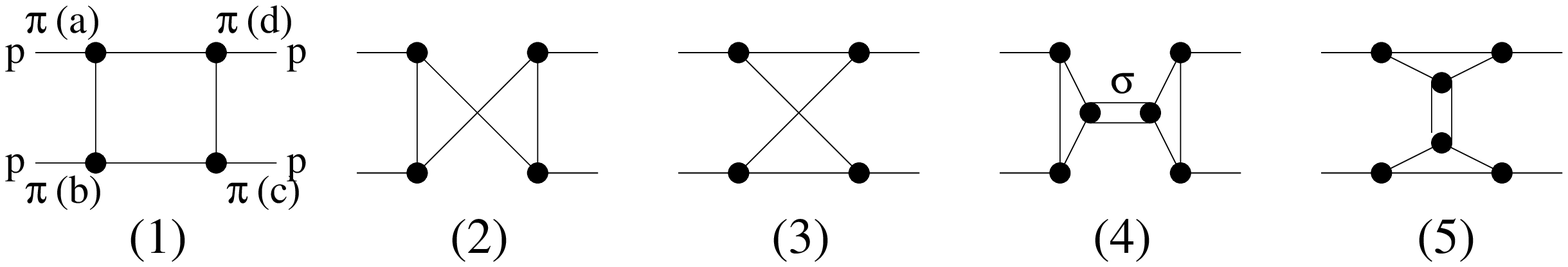,width=\linewidth}
  \caption{{\it Box and $\sigma$-meson exchange diagrams for $\pi\pi$ scattering
  (a,b,c and d are the isospin indices).}}
\end{figure}

\begin{figure}
  \begin{minipage}[b]{.45\linewidth}
    \centering\epsfig{figure=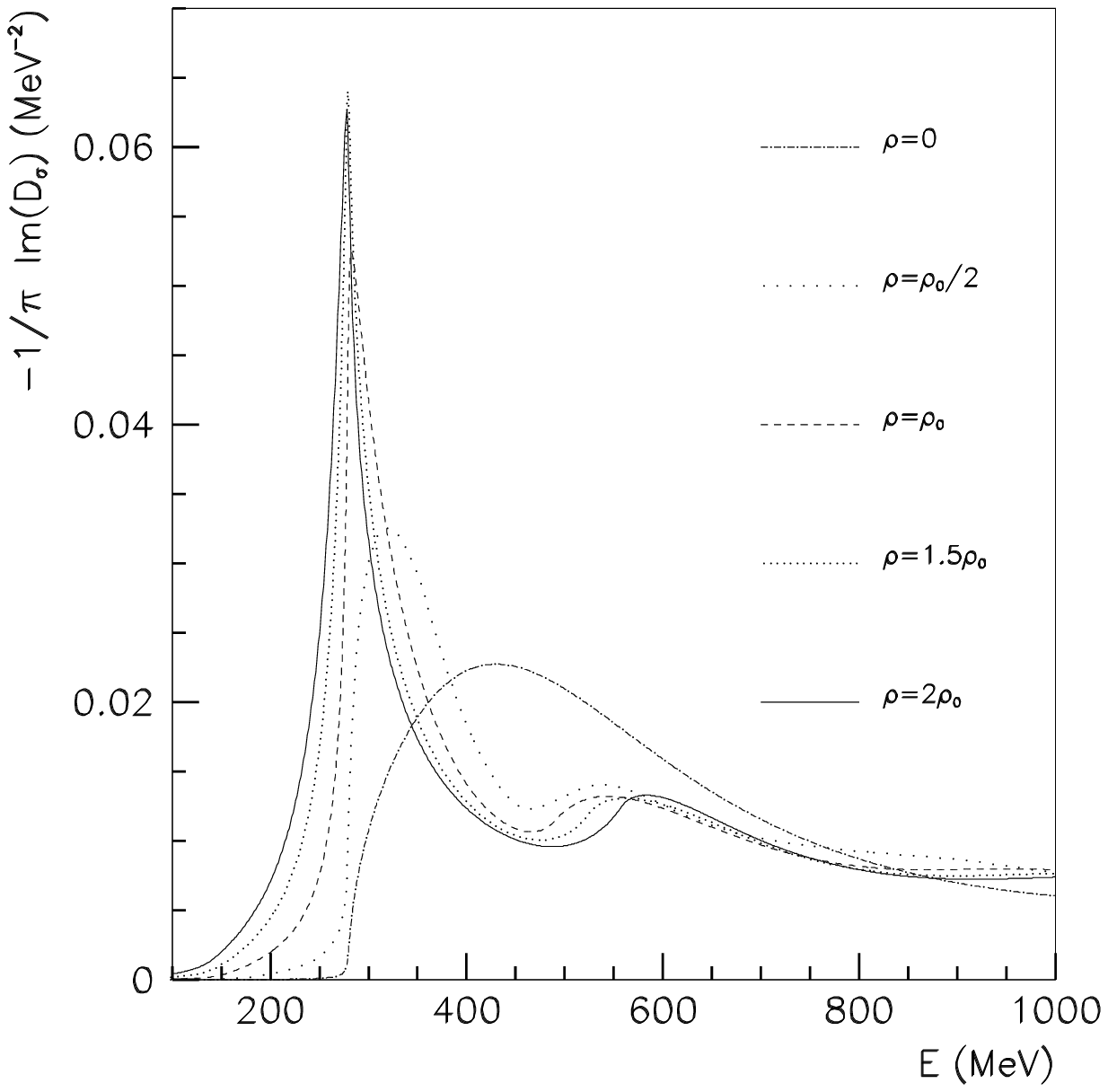,width=\linewidth}
  \end{minipage}\hfill
   \begin{minipage}[b]{0.45\linewidth}
    \centering\epsfig{figure=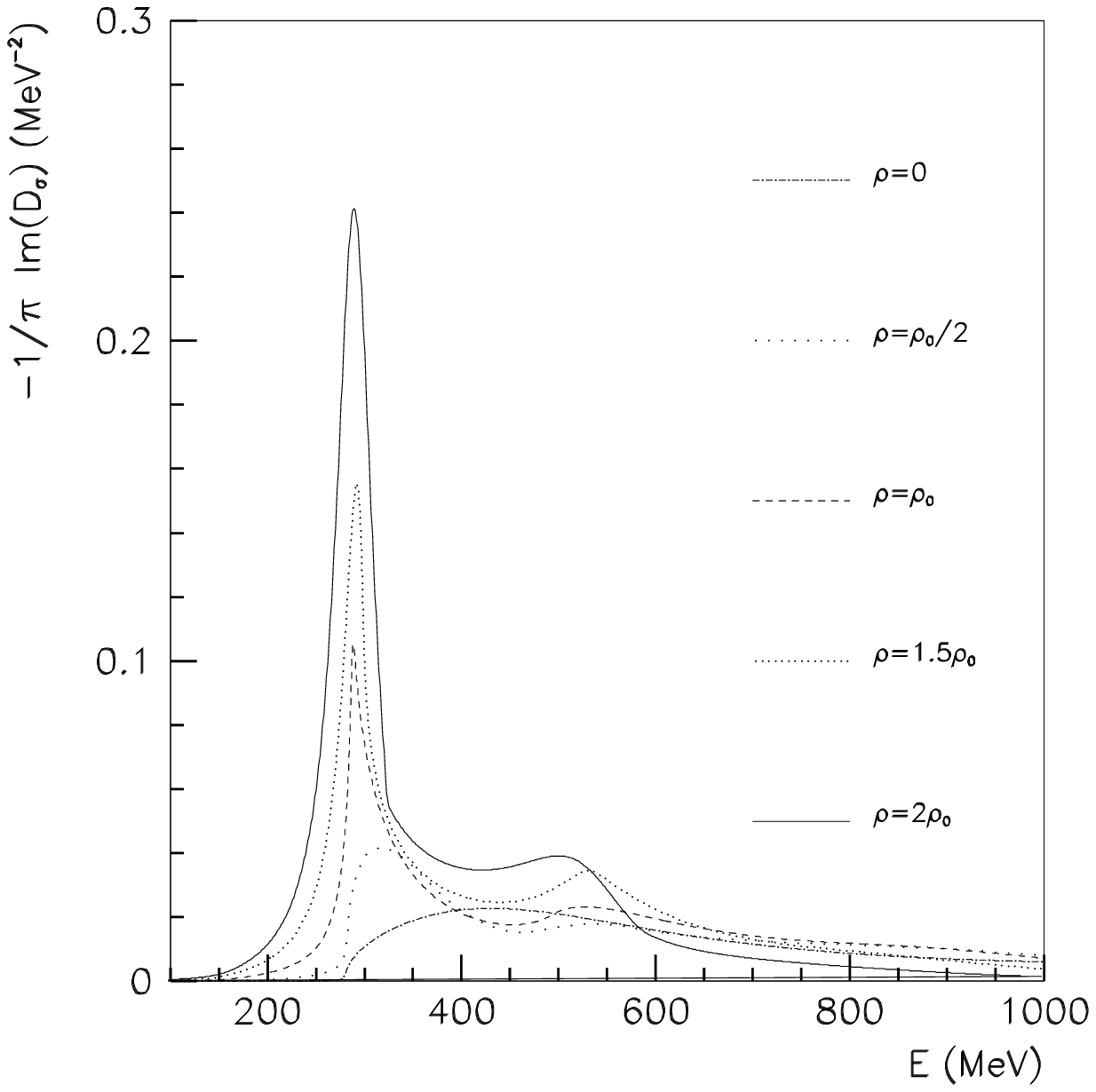,width=\linewidth}
 \end{minipage}
\caption{{\it The spectral function for the sigma meson at various densities 
($0,\, 0.5,\, 1,\,
1.5,\, 2\,  \rho/\rho_0$).  
Left~: only p-wave pionic effects. Right~: with chiral symmetry
restoration on top of p-wave effects.}}     
\end{figure}

\begin{table}                   
\caption{\it $m_\sigma$, $m_\pi$, $f_\pi$ (in MeV) , 
the scalar four-quark condensate and the 'kappa factor'
versus density. The definitions are given in the text.} 
\begin{tabular}{cccccccc}

$\rho/\rho_0$ &$m_{\sigma}$&$m_{\pi}$&$f_{\pi}$ &$\Delta Q_1$ &$\Delta Q_2$  
&$\Delta \kappa_1$  &$\Delta\kappa_2$      \\
 
\hline 
0   & 1000 & 139   & 93 &  0    & 0    & 0    & 0 \\ 
0.5 & 890  & 140.3 & 87 & -0.2  & 0.03 & 0.24 & 0.33\\ 
1   & 795  & 143.1 & 79 & -0.39 & 0.13 & 0.55 & 0.88 \\
1.5 & 649  & 148.7 & 69 & -0.57 & 0.29 & 0.96 & 1.79\\  
2   & 510  & 161   & 56 & -0.75 & 0.62 & 1.51 & 3.74 
\end{tabular} 
\label{Paramt} 
\end{table} 

\section{Discussion and Interpretation}
Four-quark condensates are fundamental quantities of non-pertubative QCD. They are known 
to play an important role in  QCD sum rule analyses of hadron spectral functions in the 
vacuum and in matter. 
As an example the density evolution of the four-quark condensate appearing in
the rho-meson case is an important input for the medium modification of this vector meson
in connection with dilepton production in relativistic heavy ion collisions. Here we 
will concentrate on to the scalar four-quark condensate $\langle(\bar q q)^2\rangle$   
and study how much it deviates from  $\langle\bar q q\rangle^2$ to asses the
evolution of quantum fluctuations with increasing density. Inserting a complete set of 
states, the scalar four-quark condensate is given as~:
\begin{equation}
\langle 0|(\bar q q)^2|0\rangle=
\langle 0|\bar q q|0\rangle \langle 0|\bar q q|0\rangle + \sum_n\,
\langle 0|\bar q q|n\rangle \langle n|\bar q q|0\rangle
\end{equation}
The first contribution to the sum are scalar-isoscalar
two-pion states, able to build a collective
sigma meson.  To evaluate (at least qualitatively) 
this quantity one can use low-energy effective theories.

\bigskip
\noindent
{\it Linear sigma model}.~ In the linear sigma model with
 chiral-symmetry breaking 
piece ${\cal L}_{\chi SB}=-f_\pi m^2_\pi \sigma$, $\langle\sigma\rangle$ plays 
the role of the condensate and $\langle\sigma^2\rangle$ relates to the 
four-quark condensate. Here we do not aim to estimate the absolute value of this 
four-quark condensate but restrict our study to its evolution with density. We thus 
define the quantity $Q_1=\langle 0|(\bar q q)^2|0\rangle(\rho)/
\langle\bar q q\rangle_{vac}^2$. Introducing the fluctuating part of the sigma field 
$s=\sigma-f_\pi$, one obtains~~:
\begin{equation}
Q_1(\rho)={\langle(\bar q q)^2\rangle(\rho)\over \langle\bar q q\rangle_{vac}^2}=
{\langle\sigma^2\rangle(\rho)\over f^2_\pi}
=1\,+\,2{\langle s\rangle(\rho)\over f_\pi}\,+\, {\langle 
s^2\rangle(\rho)\over f^2_\pi}
\end{equation}
A better measure of the fluctuations of the condensate is provided by the 'kappa factor'
defined in QCD sum rule analyses~:
\begin{equation}
\kappa_1(\rho)={\langle(\bar q q)^2\rangle(\rho)\over \langle\bar q q\rangle^2(\rho)}
=
{\langle\sigma^2\rangle(\rho)\over \langle\sigma\rangle^2(\rho)}
={1\,+\,2\langle s\rangle(\rho)/ f_\pi\,+\, \langle s^2\rangle(\rho)/ f^2_\pi\over
1\,+\,2\langle s\rangle(\rho)/ f_\pi\,+\, \langle s\rangle^2(\rho)/ f^2_\pi}
\end{equation} 
To first order in the density $\langle s\rangle(\rho)$ is governed 
by the pion-nucleon sigma term. Using a value compatible with the model 
(ignoring the pionic contribution), we take $\langle s\rangle(\rho)=-0.18\, 
\rho/\rho_0$. From a standard dispersive analysis, $\langle s^2\rangle$ 
can be expressed in terms of a phase-space integral of the sigma-meson
spectral function as~:
\begin{equation}
\langle s^2\rangle(\rho)=\int_0^{\Lambda_P}\,{d{\bf P}\over(2\pi)^3}\,
\int_0^\infty\,dE \,\left(-{1\over \pi}\right)\,Im D_\sigma (E, {\bf P})
\end{equation}
where we have introduced a momentum cutoff $\Lambda_P$ defining the range of validity 
of the effective approach. Taking $\Lambda_P\sim 1\,GeV$, and making a simple 
estimate with a sharp sigma meson of mass 1 GeV, we obtain in the vacuum a kappa 
factor of the order or larger than 2 which is very close (most probably by accident) 
to $\kappa=2.36$, generally used in the rho-meson channel. 
Finally, since we do not know the full momentum dependence of the 
sigma spectral function, we assume covariance for $D_\sigma$. We have checked that, 
using another extreme assumption (static approximation), the results are qualitatively 
similar. To reduce the uncertainty on the cutoff we prefer to present the results in 
Tabl.~1 for the quantities $\Delta Q_1(\rho)=Q_1(\rho)-(Q_1)_{vac}$ and  
$\Delta \kappa_1(\rho)=\kappa_1(\rho)-(\kappa_1)_{vac}$. For the sigma-meson propagator 
we use the form (\ref{PROP}) but with vacuum values for $m_\sigma$, $m_\pi$ and $f_\pi$, 
hence keeping only medium effects from p-wave pionic collective modes. In the actual 
calculation we have adopted the cutoff parameter $\Lambda_P=1.2\,GeV$ to (arbitrarily) 
fix the kappa factor in the vacuum to $\kappa_{vac}=2.36$ keeping in mind that 
the density evolution should not be very sensitive to this particular choice.  

\bigskip
\noindent
{\it NJL model}.~ In the NJL model the four-quark condensate can be directly calculated. 
It can be expressed in terms of a phase-space integral of the full scalar-isoscalar 
response function and subsequently in terms of the $q\bar q$ sigma meson spectral 
function~:
\begin{equation}
\kappa_2(\rho)={\langle(\bar q q)^2\rangle(\rho)\over \langle\bar q q 
\rangle^2(\rho)}=1\,+\,{1\over
f^2_\pi(\rho)}\,
\int_0^{\Lambda_P}\,{d{\bf P}\over(2\pi)^3}\,\int_0^\infty\,dE 
\,\left(-{1\over \pi}\right)\,Im D_\sigma (E, {\bf P})
\end{equation}
\begin{equation}
Q_2(\rho)={\langle(\bar q q)^2\rangle(\rho)\over \langle\bar q q)\rangle_{vac}^2}=
{\left(f^2_\pi\, m^2_\pi\right)(\rho)\over
\left(f^2_\pi\, m^2_\pi\right)_{vac}}\,\kappa_2(\rho)
\end{equation}
The sigma propagator is then unitarized by incorporating a dressed pion loop
(eq. \ref{PROP}). Notice that, in the vacuum, we exactly recover the linear sigma model
results. We show in Tab.~1 the quantities $\Delta Q_2(\rho)$ and $\Delta \kappa_2(\rho)$
which now contain the effect of chiral symmetry restoration on top of the p-wave pionic
collective modes. One sees that 
the four-quark condensate $Q_2$ increases with density when chiral
symmetry restoration is incorporated, contrary to the pure p-wave case ($Q_1$). 
We also see, by looking at the kappa factor, that chiral symmetry restoration considerably
increases the fluctuations of the condensate. This demonstrates 
that the sharp structure near $2 m_\pi$, which is mainly associated with the dropping 
sigma-meson mass, is intimately related to precritical effects with a strong enhancement 
of chiral fluctuations. In a second-order phase transition these would actually diverge in 
the chiral limit, $m_\pi\to 0$.
\vspace{0.5cm}

\vfill\eject
\noindent{\bf Acknowledgements:}

\noindent

\noindent
We wish to thank P. Schuck  for many fruitful
discussions and constant interest in this work.


\begin{thebibliography}{99}
\bibitem{Hat_99}
        T. Hatsuda, T. Kunihiro and H. Shimizu, 
        Phys. Rev. Lett. 82 (1999) 2840, 
        T. Hatsuda and T. Kunihiro,
        nucl-th/9901020, nucl-th/9902025
\bibitem{Aou_94} 
        Z. Aouissat, R. Rapp, G. Chanfray, P. Schuck and  J.
        Wambach,  
        Nucl. Phys. A 581 (1995) 471. P. Shuck {\it et al}
        contribution to these proceedings. 
\bibitem{Sch_98} 
        P. Schuck {\it et. al.},    
        nucl-th/9806069.              
\bibitem{Kre_98} 
        R. Rapp {\it et. al.}, 
        Phys. Rev. C 59 (1999) R1237    
\bibitem{Bon_96}  
        F. Bonutti {\it et al.} , the CHAOS collaboration, 
        Phys. Rev. Lett. 77 (1996) 603; N. Grion, contribution to
        these proceedings. 
\bibitem{Vic_99}
         M.J. Vicente Vacas and E. Oset,
         nucl-th/9907008. M. Vicente Vacas, contribution to these
         proceedings. 
\bibitem{Aou_99}
         Z. Aouissat, G. Chanfray, P. Schuck and  J. Wambach, 
         Phys. Rev. C61, (2000) 12202.
         P. Schuck {\it et al}, contribution to these proceedings.          
\bibitem{Da_99}
         D. Davesne, Y.J. Zhang and G. Chanfray, 
         nucl-th/9909032.         
\bibitem{Qua_95}                
        E. Quack, P. Zhuang, Y. Kalinovsky, S. P. Klevansky, 
        and J. H\"{u}fner, 
        Phys. Lett. B 348 (1995) 1. 
\end{thebibliography}
\end{document}